\documentclass[
amsmath,amssymb,
aps,
prb,
twocolumn,
superscriptaddress,%
preprintnumbers,%
byrevtex,%
floatfix%
]{revtex4-2}

\usepackage{dcolumn}
\usepackage{bm}
\usepackage{ifthen}
\usepackage[usenames]{color}
\usepackage{xcolor}
\usepackage{amssymb}
\usepackage{textcomp}
\usepackage{amsfonts}
\usepackage{amsmath}
\usepackage[]{graphicx}
\usepackage{graphpap}

\usepackage{multirow}
\usepackage{placeins}
\usepackage{float}
\usepackage{romannum}

\begin{document}

\date{\today}
\title{Coherence properties of NV-center ensembles in diamond coupled to an electron-spin bath}

\author{Reyhaneh Ghassemizadeh}
\affiliation{Fraunhofer Institute for Mechanics of Materials IWM, W\"ohlerstra{\ss}e 
11, 79108 Freiburg, Germany}

\author{Wolfgang K\"orner}
\affiliation{Fraunhofer Institute for Mechanics of Materials IWM, 
W\"ohlerstra{\ss}e 11, 79108 Freiburg, Germany}

\author{Daniel F. Urban}
\email{daniel.urban@iwm.fraunhofer.de}
\affiliation{Fraunhofer Institute for Mechanics of Materials IWM, W\"ohlerstra{\ss}e 11, 79108 Freiburg, Germany}

\author{Christian Els\"asser}
\affiliation{Fraunhofer Institute for Mechanics of Materials IWM,
W\"ohlerstra{\ss}e 11, 79108 Freiburg, Germany}%
\affiliation{University of Freiburg, Freiburg Materials Research Center (FMF), Stefan-Meier-Stra{\ss}e 21, 79104 Freiburg, Germany}

\begin{abstract}
We investigate nitrogen-vacancy center (NV) ensembles in diamond under the influence of strongly-correlated electron-spin baths. We thoroughly calculate the decoherence properties of the NV central spin for bath concentrations of $0.1-100$ ppm using the cluster-correlation expansion (CCE) method. We systematically analyze possible origins of the significant deviations in the values of the $T_2$ coherence time reported in literature. 
We demonstrate that significant variations can originate from the choice of averaging and fitting procedures used for the ensemble average and we point out the respective aspects that need to be considered, when comparing the various theoretical studies.   
Our study may ease readers to perform reliable simulations on the central spin problem. It provides an understanding and interpretation of the outcome parameters describing the dynamics of the local bath spins.   
\end{abstract}


\maketitle
\section{Introduction}

In the past two decades, solid-state based platforms for quantum technology applications like quantum sensing and quantum computing became increasingly popular and promising due to their outstanding properties like, e.g., their long spin coherency even at room temperature~\cite{nay11,bal09,bargill13,sae13,wolf21,becher23,jelezko04,cai13}. 
The negatively charged nitrogen vacancy (NV) defect complex in diamond, which consists of a substitutional
nitrogen atom with a vacant nearest-neighbor carbon site and an additional electron, is one of the candidates for the near future qubits due to its extraordinary spin properties~\cite{doh12,bern17}.
The electronic ground state of the NV center is a triplet state, $S = 1$, with a spin life-time of about 1 ms at  room temperature~\cite{gali19, zhao12}. However, this long spin coherency can be limited by the interaction of the NV electron spin with surrounding spins and crystal defects. 
In highly purified diamond crystals, the main source of the NV-spin decoherency is known to be nuclear spins of surrounding $^{13}$C isotopes~\cite{wyk97,bauch20}. This decoherence mechanism is well understood within the quantum mechanical framework~\cite{hall14,zhao12,maze08,stan10,shi14,witzel06,yang17}. 
However, during the growth of diamond by chemical vapor deposition (CVD) or ion implementation for creating the NV centers, paramagnetic impurities, e.g., substitutional $^{14}$N (so-called P1 centers), neutral NV$^0$, NVH$^-$, or carbon-vacancy pairs VV$^0$, are frequently formed~\cite{rod21,shinei22,nunn22, ferr22,barry20,li21}. These defects have unpaired electron spins, which can strongly couple to the NV center and play a crucial role in decreasing the spin coherency and, hence, the efficiency of the NV-based devices. Consequently, many experimental~\cite{bargill13,step16,rod21,abey16,barry20,shinei22,bauch20,ed21,hay20,li21} and theoretical~\cite{witzel12,wang13,bauch20,park22,hahn24,marcks24} studies investigated the decoherence of a central spin in a strongly coupled spin environment in the past decade.

\begin{figure}[]
\begin{center}
	\includegraphics[width=0.9\columnwidth]{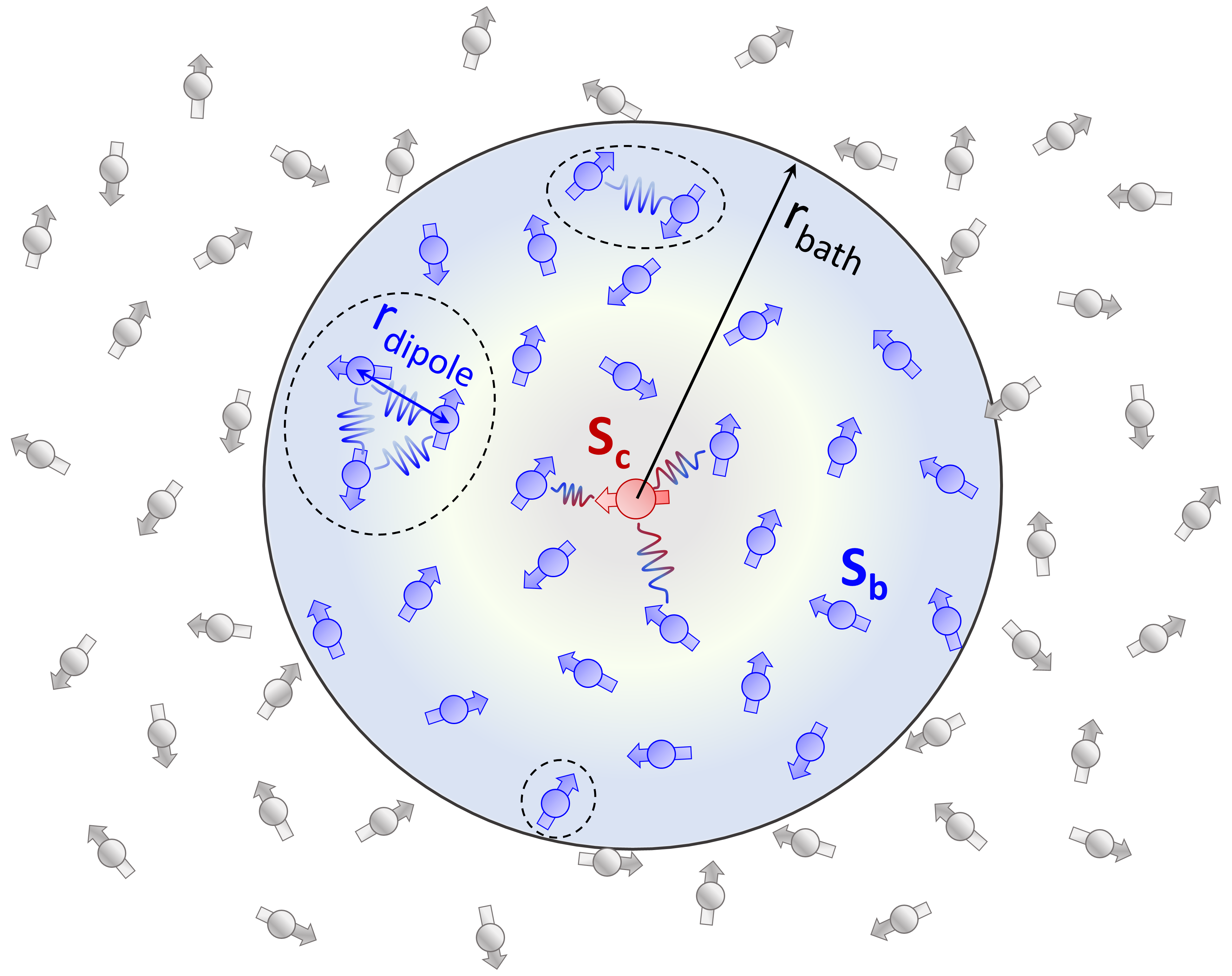}
\caption{Schematic of the central spin problem. The central spin (S$_c$) interacts with the bath spins (S$_b$) within the radius of r$_{bath}$. Within the CCE approximation, bath spins which are closer than r$_{dipole}$ are grouped into clusters of order $n=1,2,3,...$ as exemplary shown by the dashed lines.}\label{fig:spin-bath}
\end{center}
\end{figure}

In a central spin problem, the schematic of which is shown in Fig.\ref{fig:spin-bath}, the total Hamiltonian is written as~\cite{oni21}
\begin{equation}
\label{eq:hamiltonian}
\hat{H} = \hat{H}_S + \hat{H}_{SB} + \hat{H}_B, 
\end{equation}
where $\hat{H}_S$ is the Hamiltonian of the central spin, $\hat{H}_{SB}$ accounts for the interaction between the central spin and the bath spins and $\hat{H}_B$ denotes the Hamiltonian of the bath spins. More detailed explanations can be found in Sec.~\ref{sec:cce_model} and the Refs.~\cite{hall14,oni21,park22,hahn24,maile24} 

Generally, the normalized coherence decay signal $\mathcal{L}(t)$ of a central spin in an interacting spin bath is well described by classical stochastic noise, e.g., the Ornstein-Uhlenbeck process~\cite{lang10,witzel14, abey16}. It can be fitted by an exponential decay function 
\begin{equation}
\label{eq:exp}
\mathcal{L}_{\rm fit}(t)=\exp\left[-\left(\frac{t}{T_2}\right)^p\right], 
\end{equation}
with the stretched exponential parameter $p$ and the coherence time $T_2$ as fit parameters~\cite{lang10,abey16,bauch20}.  These contain important information on the NV center spin decoherency. Here, $p$ determines the curvature and shape of the decay and $T_2$ quantifies the relevant time scale for the decay. The spin-echo decoherence signal of a single NV center in a weakly coupled bath of nuclear spins is characterized by a decay with $p\sim2$--3~\cite{lang10,witzel14,abey16,maze08}. On the other hand, for an ensemble of NV centers in a strongly correlated electron spin bath, the decay signal can be well approximated by Eq.\ (\ref{eq:exp}) with $p \approx$  1--1.5~\cite{hanson08,park22,bauch20,li21,shinei22}.
Recently it has been reported, that $p$ can even take values below 1 in the case of a quasi two-dimensional spin environment \cite{davis23,hahn24}. 
Moreover, for the case of strongly coupled P1 electron spins, an almost linear dependency of $T_2$ on the inverse of the concentration of P1 centers ($\rho_{P1}$) has been reported from both, theoretical calculations~\cite{park22,bauch20,wang13,hahn24,marcks24} and experimental measurements~\cite{step16,bauch20}. However, the reported $p$ and $T_2$ values vary over a wide range, as demonstrated by the compilation in  Tab.~\ref{tab:compares-ps}.

The broad range of $T_2$ and $p$ values obtained in different theoretical studies motivates us to analyze the possible sources of this variety. Therefore, we thoroughly analyze possible origins for the large variation of about 20\% and 30\% in $T_2$ and $p$ values, by considering different analytical and numerical criteria.  
We use the established CCE method to calculate the Hahn-echo~\cite{witzel06} decoherence signal of ensembles of NV centers in various concentrations of the electron-spin bath ($\rho_e= 0.1-100$ ppm). We chose the pure electron bath environment because the existence of defects with an extra electron have as an important impact on the NV center decoherence as the P1 centers~\cite{park22}, and there is still a lack of studies in this regard in the literature. 
Moreover, the conceptional sources of discrepancy in the $T_2$ and $p$ values are independent of the addition of hyperfine-field (HF) coupling between nuclear and electronic bath spins.

By varying the spatial configuration of the spins surrounding the NV center we obtained a wide range of coherence decay behaviors for individual bath configurations, from an extremely fast decay to a slow decay. This leads to a wide range of values of T$_2$ and $p$ for each individual spatial spin configuration at given $\rho_e$ with a large standard deviation. Hence, including enough realizations in the ensemble average is crucial to obtain a robust and reliable estimate of the coherence time $T_2$. Moreover, the convergence parameters of the CCE method and the fit uncertainties in numerical fits of the decay function of Eq~\ref{eq:exp} are further origins of the inconsistency in the $T_2$ and $p$ values even at the same level of theory. 

The manuscript is organized as follows. As a starting point, in Sec.~\ref{sec:lit} we briefly review the available literature on theoretical studies of the decoherence of a NV spin subject to an environmental spin bath. In Sec.~\ref{sec:cce_model} we outline our theoretical model and the numerical settings used in the CCE calculations. Our results and the related discussions in Sec.~\ref{sec:results} are structured by first analysing the limitations in the ensemble average in Sec.~\ref{subsec:ensemble_average} and then discussing the relevance of the order of the CCE approximation in Sec.~\ref{subsec:order}. Our results for the decoherence signal of an ensemble of NV centers in various concentrations of the electron-spin bath ($\rho_e= 0.1-100$ ppm) are presented in Sec.~\ref{subsec:coherency_ppms} and augmented by a detailed analysis of different fitting strategies in Sec.~\ref{subsec:fit_analysis}. We summarize our findings in Sec.~\ref{sec:summary}.

\section{Review of literature results}\label{sec:lit}

\begin{table}[]
	\vspace{0.2cm}
	\begin{tabular}{l c c c c c c c c c}
	\hline 
	 &  \multicolumn{2}{c}{1 [ppm]}&& \multicolumn{2}{c}{10 [ppm]} && \multicolumn{2}{c}{100 [ppm]}\\
   	 & $T_2$ [$\mu$s] & $p$  && $T_2$ [$\mu$s]  & $p$  && $T_2$ [$\mu$s] & $p$ \\
	\hline 
	Ref.\cite{hahn24}- HF(r)   & 225  & 1.03    && -    & -    && -       & -  \\
	Ref.\cite{marcks24}- HF(r) & 230  & 1.2-1.3 && 25   & 1.2-1.3 && 3.5  & 1.2-1.3 \\
	Ref.\cite{park22}- HF(r)   & 407  & 0.8     && 39   & 0.9  && 3.3     & 0.8 \\
	Ref.\cite{bauch20}- HF(r)  & 70-140 & 1.5   && 8-10 & 1.5  && 0.6-0.8 & 1.5 \\
	Ref.\cite{park22}- HF(a)   & 250  & 0.9     && 25   & 1    && 2.5     & 0.8 \\
	Ref.\cite{hahn24}    & 49.2 & 1.43    && -    & -    && -       & -   \\
	Ref.\cite{park22}    & 91.2 & 1.1     && 8.9  & 1.1  && 0.87   & 1.1 \\
	Ref.\cite{wang13}$^{*}$ & 20 & 2-3    && 2.0  & 2-3  && 0.2    & 2-3 \\
	This work            & 52.7 & 1.44    && 5.4  & 1.4  && 0.53   & 1.5 \\
	\hline  
	\end{tabular}
\caption{Theoretical values of $T_2$ and $p$ for the coherence decay of the NV ensemble in three dimensional P1 spin baths with concentrations of 1, 10, and 100 ppm, as reported in literature. Some calculations include hyperfine interaction between electron and nuclear spin and methods use random or aligned Jahn-Teller axes, denoted as HF(r) and HF(a), respectively. The others consider a pure electron spin bath. $^{*}$No changes in values with or without HF are reported in Ref.~\cite{wang13}. }
\label{tab:compares-ps}
\end{table}

Table~\ref{tab:compares-ps} summarizes the theoretical results for $T_2$ and $p$ for three dimensional NV center ensembles in diamond in different concentrations of P1 centers, including nuclear and electron spin baths, as reported in literature to the best of our knowledge. In a P1-center spin-bath the hyperfine interaction (HF) between the electron and the respective nuclei spin gives rise to an additional term in the Hamiltonian of the bath which is not present in the case of a pure electron-spin bath. Moreover, the splitting of the hyperfine levels of the P1 bath is different based on the orientation of the Jahn-Teller axes of the P1 centers~\cite{park22}.

In an earlier simulation of Wang et al.~\cite{wang13} a so-called $P$-representation 
method~\cite{zhang07} was used to calculate the decoherence of the NV spin coupled to the surrounding P1 spins. Although the derived linear dependency of the $T_2$ time on the inverse of the P1 center concentration agrees to the more recent available studies, the values of $T_2$ are highly underestimated compared to the later works. Moreover, Ref.~\cite{wang13} reports no significant difference in the $T_2$ times with and without considering the HF coupling of the P1 centers. 

The more recent work of Bauch et al.~\cite{bauch20} employed a semi-classical approach to estimate the values of $T_2$ as function of the concentration of P1 centers, $\rho_{P1}$. 
Here, P1 centers are found to be the dominating source of the NV decoherence for $\rho_{P1} \geq$ 0.5 ppm where $T_2\propto(\rho_{P1})^{-1.07}$. Including HF coupling with a random distribution of JT axes in their theoretical model, they obtained $T_2$ values which are considerably below those obtained later by quantum mechanical approaches~\cite{park22,hahn24,marcks24}. They argue that their semi-classical approach only poorly captures the dynamics of the bath electron spins. Furthermore, the inaccuracy of the employed fitting caused a broad spread of their computed values.

In a recent work, Park et al.~\cite{park22} employ the quantum mechanical approach of cluster correlation expansion (CCE)~\cite{witzel06,yang08,yang09,zhao12,witzel14,zhang20} to accurately determine the dynamics of strongly correlated spin baths. They considered the cases where the HF coupling is included with either random or aligned JT axes, or it is totally excluded, which mimics the behavior of a pure electron-spin bath. They show that including HF interactions strongly suppresses the electron-spin flip-flop transitions of the NV center and leads to approximately 4.5 times larger $T_2$ values compared to the experimental values. The addition of electron spins to the bath with at least the same concentration as P1 centers decreases their $T_2$ values~\cite{park22} to the experimental values reported by Bauch et al.~\cite{bauch20}. 

In the very recent studies of Sch\"atzle et al.~\cite{hahn24} and Marcks et al.~\cite{marcks24} the influence of the diamond layer thickness on the coherency of the NV ensembles in the P1 spin baths is investigated using the CCE method. Both studies indicate an increase in spin coherency for thin films of diamond. Their reported T$_2$ values for bulk diamond vary within less than 3\% in comparison with each other but they deviate by about 50\% from the CCE results of Park et al.~\cite{park22}.

\section{Theoretical Model}\label{sec:cce_model}
We use the generalized cluster correlation expansion method (gCCE)~\cite{oni21,oni20} as implemented in the python module PyCCE~\cite{oni21} to calculate the coherence function of the central electron spin of the NV center in a bath of paramagnetic spins. The components of the Hamiltonian for the system based on Eq.~(\ref{eq:hamiltonian}) are:
\begin{equation}
\begin{split}
\label{eq:Hamiltonian_components}
 &\hat{H}_S  = \mathbf{SDS} + \mathbf{B}\gamma_{S}\mathbf{S}, \\
 &\hat{H}_{SB} =   \sum_{i} \mathbf {S} \mathbf{A}_i \mathbf{I}_i \\
 &\hat{H}_B = \sum_{i>j} \mathbf{I}_i \mathbf{K}_{ij} \mathbf{I}_j + \mathbf{B} \gamma_i \mathbf{I}_i
\end{split}
\end{equation}
Here, $\mathbf{S}$ ($\mathbf{I}$) is the central (bath) spin operator, $\gamma_{S/i}=\gamma_e$ is the gyromagnetic ratio of the central (bath) electron spin, $\mathbf{D}$ is the zero-field splitting tensor of the NV center, and $\mathbf{B}$ is the external magnetic field.
The coupling tensor $\mathbf{K}$ between the bath spins is approximated from the magnetic point dipole-dipole interactions as
\begin{equation}
\begin{split}
\label{eq:dipolar_coupling}
\mathbf{K}_{ij} = -{\gamma_e}^2 \frac{\mu_0\hbar^2}{4\pi} \left[ \frac{3\vec{r}_{ij}\otimes \vec{r}_{ij}  - |r_{ij}|^2 \mathbf{I}}{|r_{ij}|^5}\right]
\end{split}
\end{equation}
with $\vec{r}_{ij}$ being the distance vector between the two spins and  $\mathbf{I}$ is the 3$\times$3 identity matrix.
The coupling $\mathbf{A}$ between the NV center and the bath spins is of the same form.

The CCE method quantum mechanically describes the decoherence of a central spin (or qubit) in a bath of surrounding spins by capturing the time evolution of the bath dynamics. In this approach, the surrounding bath spins are grouped into clusters of $n$ spins (n = 1, 2, etc.) in terms of the strength of their spin-spin dipolar interaction. The contribution of each spin cluster is calculated quantum mechanically. The total coherence function $\mathcal{L}(t)$ of the central spin is factorized into sets of irreducible contributions from spin bath clusters~\cite{yang08,yang09}. 
The validity of the CCE method for solving the problem of the central spin in spin baths of both nuclei~\cite{witzel06,zhao12,seo16,ye19,oni20,oni21,sajid22,oni23} and electrons~\cite{witzel10,witzel12,hay20,park22,hahn24} has been well established. However, in case of more strongly correlated systems, like electron spin baths, more strict convergence criteria need to be applied. The important convergence parameters in CCE are: (\romannum{1}) the CCE \textit{order} which defines the maximum number of spins interacting within a cluster, (\romannum{2}) the maximum distance $r_{\rm dipole}$ between two bath spins that can be assigned to the same cluster, and (\romannum{3}) the cut-off radius $r_{\rm bath}$ which determines the spherical region around the central spin in which bath spins have an active influence on the central spin decoherency in the considered timescale(see Fig.~\ref{fig:spin-bath}).

Our model systems are constructed as follows. We start from a cubic diamond-crystal unit-cell with a lattice constant of 3.57 {\AA} and expand it to a cubic supercell with an edge length of approximately 4000 {\AA}. The NV center is placed at the center of this supercell with the $z$ axis set to be the trigonal symmetry axis of the NV center along the [111] direction of the diamond.
The electron spins of the bath, with spin $I=1/2$ and gyro-magnetic ratio $\gamma_e$ = -17608.597 rad~ms$^{-1}~G^{-1}$, are randomly distributed at carbon sites in diamond with concentrations of $\rho_{e} = 0.1 - 100$ ppm. An external magnetic field of $B = 100$ G is applied along the $z$ axis. We examined the changes in coherence time for different values of the magnetic field. Like Bauch et al.~\cite{bauch20} we found that the NV center coherency is insensitive to the value of the magnetic field if $B > 10^{-3}$ G for low concentrations of $\rho_{e}$ (e.g., 0.1 ppm) and $B > 1$ G for higher concentration (e.g., $\rho_{e}$ = 5 ppm). 

The coherency function $\mathcal{L}(t)$ is calculated for the loss of the relative phases of the qubit levels $m_{s}=0$ and $m_s =+1$ in presence of a Hahn-echo $\pi$-pulse series after a free evolution time  $t=2\tau$. Here, $\tau$ is the time between successive $\pi$ pulses. 

High-order CCE calculations for highly correlated bath spins, such as electrons, suffer from the problem of coherence function divergence at certain frequencies \cite{witzel12}. This is due to the suppression of the flip-flop dynamics by the magnetic field gradients generated by the bath spins outside the clusters \cite{witzel12}. To avoid this problem, the contribution of all other bath spins outside a cluster is taken into account as a mean-field effect via Monte Carlo sampling over 80-300 random initial spin states of the bath \cite{oni21,oni20}.
Finally, the ensemble-averaged coherence function is calculated by averaging over 500 spatially different configurations of bath spins.

\begin{figure}[]
\begin{center}
	\includegraphics[width=\columnwidth]{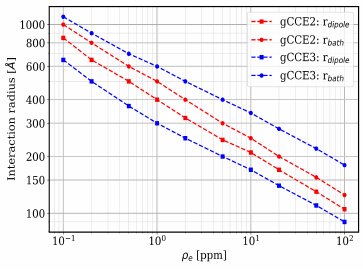}
\caption{Converged values of $r_{\rm dipole}$ and $r_{\rm bath}$ with respect to $\rho_e$ for the 2$^{nd}$ and 3$^{rd}$ cluster orders, abbreviated as gCCE2 and gCCE3, respectively.}\label{fig:cce_convergance}
\end{center}
\end{figure}

We performed convergence tests for the choice of parameters $r_{\rm dipole}$ and $r_{\rm bath}$ for both, the 2$^{nd}$ and the 3$^{rd}$ order cluster approximations (gCCE2 and gCCE3). We consider the parameters to be converged  when the variations of the extracted fitting parameters T$_2$ and $p$ evaluated from 60\% of the decay curve are less than 5$\%$. The converged values of $r_{\rm dipole}$ and $r_{\rm bath}$ as function of the electron-spin density $\rho_e$ are displayed in Fig.~\ref{fig:cce_convergance} and selected values are summarized in Tab.~\ref{tab:conv-bath}.
Regarding gCCE2, we obtain the approximate relation $r_{\rm bath}\approx 1.2 r_{\rm dipole}$ for all the considered bath spin concentrations in agreement with previous works (see e.g.~the SI of Ref.~[\onlinecite{park22}]), while in the case of gCCE3, the relation \mbox{$r_{\rm bath}\approx 2 r_{\rm dipole}$} holds for the different $\rho_e$.

\begin{table}[]
	\begin{tabular}{l l c c c c }
	&$\rho_e$          & 0.1 ppm & 1 ppm & 10 ppm & 100 ppm\\
	\hline  
	gCCE2&$r_{\rm dipole}$ [nm]  & 850 & 400 & 210 & 105 \\ 
	    &$r_{\rm bath}$ [nm]    &1000 &500 & 250 & 125 \\ 
	    &$\bar{N}_b$ ($\sigma_{\bar{N}_b})$  &74 (8) &92 (9) & 117 (10) & 145 (12)\\
	\hline  
	gCCE3&$r_{\rm dipole}$ [nm]  & 650 &  300 & 170  & 90 \\ 
	    &$r_{\rm bath}$ [nm]    & 1100  &  600 & 340   & 180  \\ 
	    &$\bar{N}_b$ ($\sigma_{\bar{N}_b})$  &99 (9) & 141 (6) & 290 (16) & 423 (21)
	\end{tabular}
\caption{The converged values of $r_{\rm dipole}$ and $r_{\rm bath}$, given in {\AA}, for various bath spin concentrations ($\rho_e$) in ppm, the corresponding averaged number of bath spins, ($\bar{N}_b$), and their standard deviation ($\sigma$).}
\label{tab:conv-bath}
\end{table}

\section{Results \& Discussion }\label{sec:results}

\subsection{Ensemble average over spin positions}
\label{subsec:ensemble_average}
The decoherence behavior of an individual NV center is strongly dependent on the specific spatial distribution of the surrounding bath spins \cite{wang13,maze08,hay20}. In order to obtain the characteristic time $T_2$ describing an ensemble of independent NV centers, it is therefore essential to average over a sufficiently large number of bath configurations. In the literature, ensembles of $80$ to $10^4$ individual NV+bath configurations were considered using different theoretical approaches \cite{wang13,bauch20,park22,marcks24}. 
In this section, we analyze the effect of using a finite-size set of configurations and show how many bath configurations are needed for an average that yields sufficiently small numerical uncertainties in the extracted $T_2$ and $p$ values. 
To achieve this, we first compute the decoherence for $N_{\rm max}=500$ random spatial configurations for each spin bath concentration. Then, we evaluate the averaged decoherency for ensembles of $N$ = 50 -- 250 bath configurations, picked randomly from the full set of 500 configurations, and repeat this 200 times for each $N$. For each of these ensemble averages we obtain the $T_2$ and the exponential power $p$ from a fitted exponential function using Eq.~(\ref{eq:exp}). By this procedure, we obtain a histogram of $T_2^{(N)}$ and $p^{(N)}$ values that may be obtained when conducting the ensemble average with $N$ configurations, compared to using the full set of $N_{\rm max}$, at given spin concentration $\rho_e$.

\begin{figure}[]
\begin{center}
	\includegraphics[width=\columnwidth]{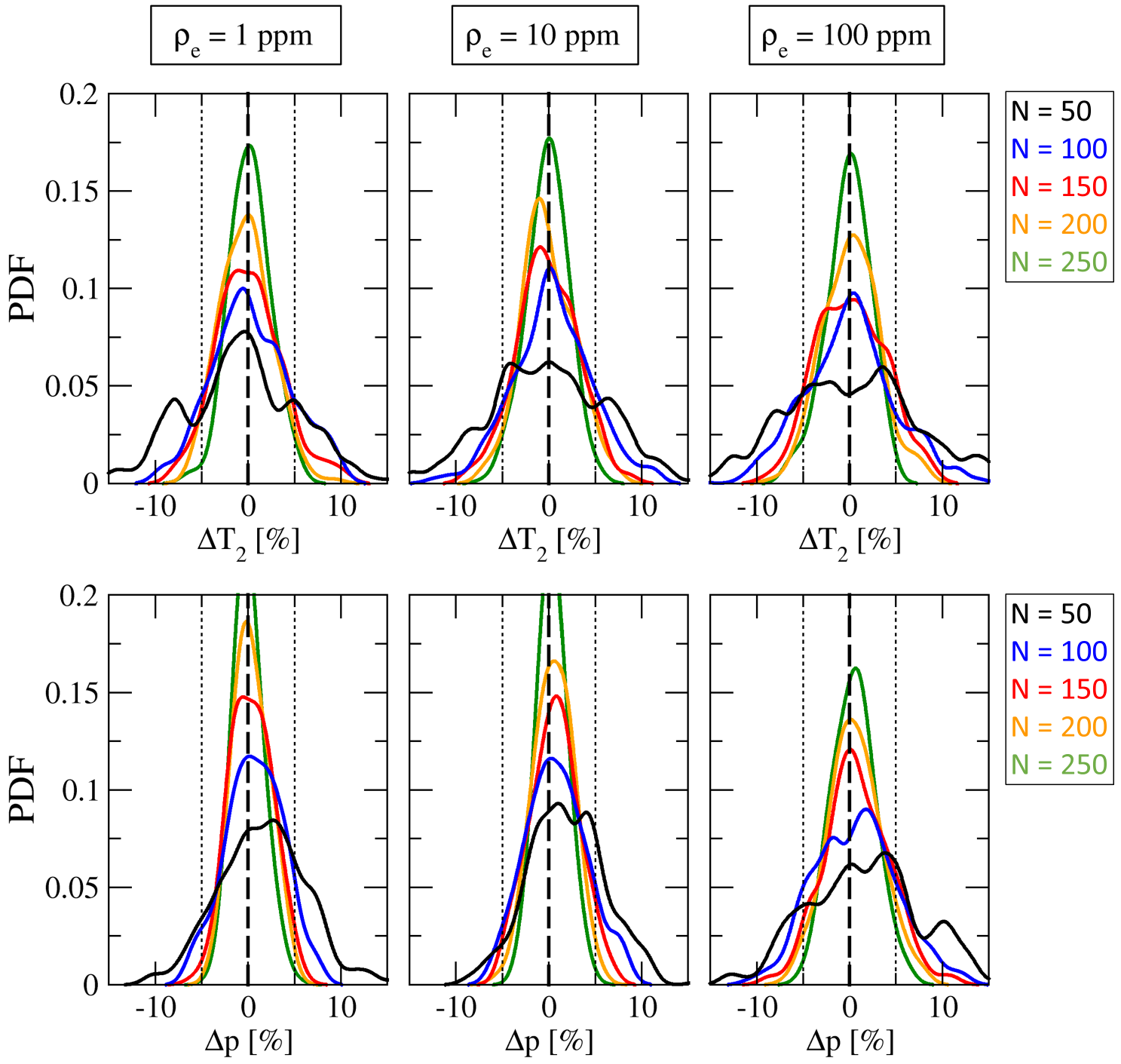}
\caption{Relative variation in $T_2$ time (top) and $p$ values (bottom) when averaging over N = 50--250 spatial configurations of the spin bath with concentration 1 ppm (left), 10 ppm (middle), and 100 ppm (right). Each panel  shows the probability distribution function (PDF) of errors relative to an ensemble average obtained from 500 configurations (dashed lines).}\label{fig:ensemble_average}
\end{center}
\end{figure}

We plot the results of this analysis via the probability distribution function (PDF) of the relative deviations, 
$\Delta T_2 = \left(T_2^{(N)}-T_2^{(N_{\rm max})}\right)/T_2^{(N_{\rm max})}$ and 
$\Delta p = \left(p^{(N)}-p^{(N_{\rm max})}\right)/p^{(N_{\rm max})}$,  in Fig.~\ref{fig:ensemble_average}. 
This illustrates the spread of values and the uncertainty that statistically can be obtained when taking an ensemble average of finite size $N$. 
Increasing the number of configurations that enter the ensemble average narrows down the spread of $T_2$ and $p$ values, and we find that an averaging over 250 samples is sufficient to yield a maximum uncertainty of approximately $\pm$ 5\%. 

\subsection{Order of the clusters expansion}\label{subsec:order}

\begin{figure}[h!]
\begin{center}
{\includegraphics{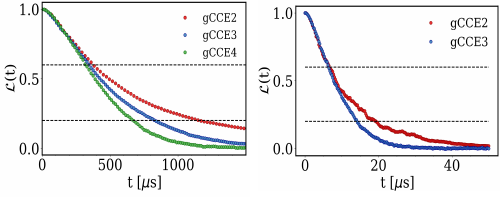}}
\put(-205,81){(a)}
\put(-86,80){(b)}  
\caption{Decay of the coherence function $\mathcal{L}(t)$ of a NV center in diamond subject to an electron spin bath for different orders of the CCE approximation. The ensemble average is obtained from 250 individual spatial configurations of electron spins with a concentration of (a) $\rho_e =$ 0.1 ppm and (b)  $\rho_e =$ 5 ppm. Dashed lines indicate different ranges of the decay as analyzed in Sec.\ \ref{subsec:fit_analysis}.
}\label{fig:cce_orders}
\end{center}
\end{figure}

Higher order cluster approximations of CCE are becoming computationally demanding, especially for higher concentrations of bath spins. For the weakly interacting spin baths of nuclei the use of the second order cluster approximation in CCE calculations has been well established~\cite{oni20,oni21,sajid22,oni23}. Also in the case of strongly interacting spin baths of electrons it has been shown that the second order cluster approximation can be sufficient for isotope-enriched silicon~\cite{witzel10} and for NV centers surrounded by very high concentrations (60-200 ppm)~\cite{witzel12} of P1 centers.
A detailed study of Witzel et al.~\cite{witzel12,witzel14} on the 2$^{nd}$, 3$^{rd}$ and 4$^{rd}$  order cluster approximations for baths of electron spins indicates that although the second order (CCE2) result for the averaged decay differs from the third order (CCE3) result, in the range of short times the decay appears to be captured well by CCE2 approximation. Park et al.~\cite{park22} stated that the CCE2 approximation is sufficient for their CCE simulations. However, the recent study of Marcks et al.~\cite{marcks24} uses CCE4 for the same system and with significantly different result, cf. Tab.~\ref{tab:compares-ps}.

We performed gCCE calculations for the 2$^{nd}$, 3$^{rd}$ and 4$^{th}$ order of spin clusters for low concentrations of 0.1-0.2 ppm. For higher concentrations up to 100 ppm we restricted our investigation to  gCCE2 and gCCE3 due to the high computational cost of gCCE4 at these high spin concentrations. For individual bath configurations it may happen that the coherence function $\mathcal{L}(t)$ obtained by the gCCE2 and gCCE3 approximations are numerically almost indistinguishable. However, in our results for an ensemble average over 250 different spatially distinct bath configurations we always see a difference in their decay behavior for times of the order of $T_2$ or larger, as illustrated in Fig.~\ref{fig:cce_orders}.

As already pointed out by Witzel et al.~\cite{witzel14}, the criterion of convergence depends on which range of the decay time is taken into account (e.g., for fitting the noise model). On shorter times the results of CCE2 and CCE3 may be considered as converged. On the other hand, when considering longer times on the order of $T_2$ or more, the CCE2 overestimates the coherence time roughly by a factor of 1.2 and 1.4 compared to the CCE3 and CCE4, respectively. 

In the following, the importance of the choice of higher cluster order to extract more stable T$_2$ and $p$ values  by fitting the analytic function of Eq.~\ref{eq:exp} to the numerical CCE data is demonstrated by our results.

\subsection{Coherence of NV ensembles}
\label{subsec:coherency_ppms}

\begin{figure}[h!]
\begin{center}
\includegraphics[width=\columnwidth]{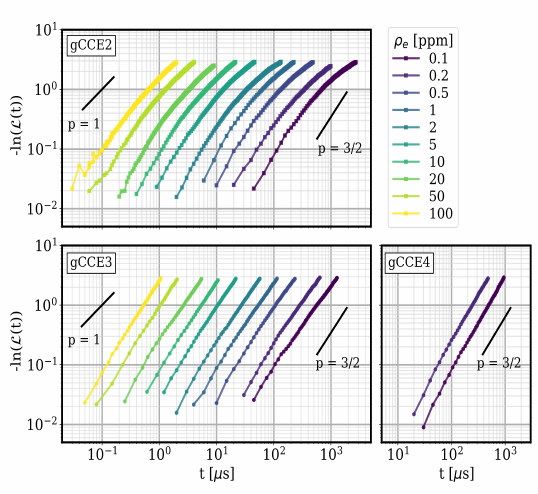}
\caption{Coherence decay of an NV center in electron-spin baths of $\rho_e =$ 0.1 -- 100 ppm plotted in double-logarithmic graphs for three different cluster orders, gCCE2, gCCE3, and gCCE4. The ensemble averages were obtained from 500 individual spatial spin configurations at each concentration. Short solid black lines indicate the slopes for selected $p$ values, for comparison. 
}\label{fig:decay_ppm}
\end{center}
\end{figure}

We evaluated the coherence decay $\mathcal{L}(t)$ for an ensemble average of NV center spins 
obtained by gCCE2 and gCCE3 calculations for electron-spin baths of $0.1 - 100$ ppm concentration, and gCCE4 calculations  for the most dilute spin baths of 0.1--0.2 ppm.
In order to display and compare the curves, we plot $-\ln (\mathcal{L}(t))$ vs.~$t$ in a double-logarithmic graph in Fig.~\ref{fig:decay_ppm}. 
In the double-logarithmic graph the slope of the curve determines the (local) value of $p$, since there is in general not a single value of $p$ which describes the entire curves, i.e.\ the full time span~\cite{hall14,davis23,hahn24}. 
However, as shown in Fig.~\ref{fig:decay_ppm}, using the higher order clusters in the CCE calculations leads to a nearly perfect linear behavior of the $-ln (\mathcal{L}(t))$ vs. $t$ in logarithmic scale for the examined time ranges shown. 

The value of $p$ will depend on several aspects of the physical system under investigation, most importantly the types of interactions that are considered and also the dimensionality of the system. Several theoretical studies have derived a certain $p$ value for a specific setting, see e.g. Refs.\ \cite{bauch20,hall14}. Moreover, depending on the system under investigation, there can be different parameter regimes defined by the relative interaction strengths or other relevant energy scales that yields a cross-over between different $p$ values for different time regimes \cite{hall14,davis23,hahn24}.

For the gCCE2 calculations, we observe a decoherence behavior on short times characterized by a slope of $p\approx1.5$ for lower concentrations and $p\approx1$ for very high concentrations of $\rho_e$ = 50 -- 100 ppm (c.f.~short black lines in Fig.~\ref{fig:decay_ppm}) 
On the other hand, for the gCCE3 calculations, the slope of the linear fit is approximately 1.5, regardless of the bath spin concentrations. The same is true within gCCE4 for $\rho_e = 0.1$ and 0.2 for which we observe an almost perfect linear relationship over the whole time range. 

We emphasize that the more linear the relationship of numerical data for $-\ln (\mathcal{L}(t))$ vs.\ $t$ in a log-log-plot gets, the more accurate a numerical fit using function (\ref{eq:exp}) becomes. On the other hand, any deviation from this linear relationship indicates, that the results obtained by fitting
will depend on the choice of fitting procedure and data range. Note that this does not only apply to computational data obtained under the constraint of theoretical approximations and the choice of numerical convergence criteria but as well to experimental data.
We analyse and quantify this dependence in detail in the following subsection.

\begin{figure*}[]
\begin{center}
{\includegraphics{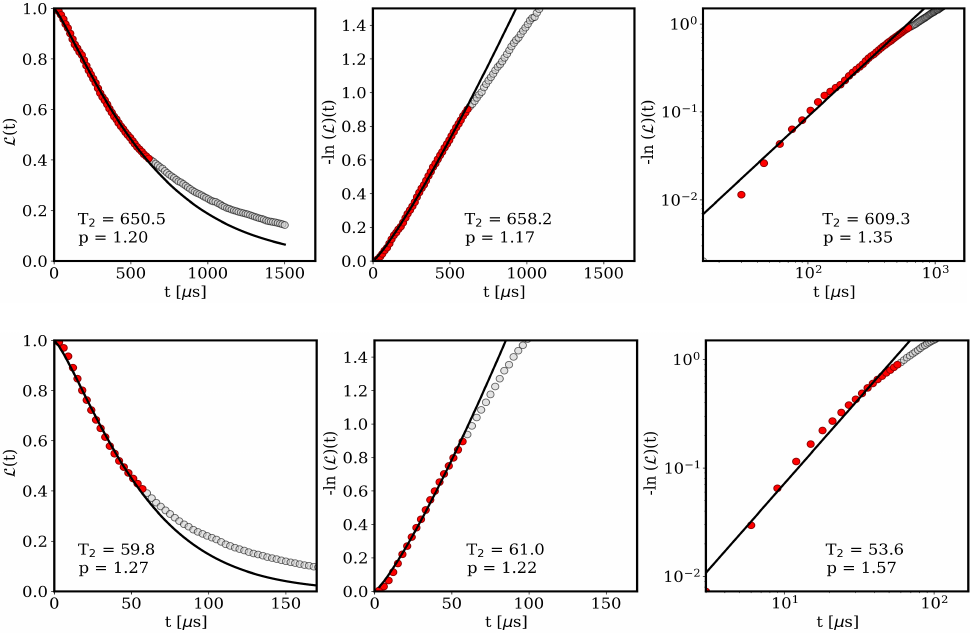}}
\put(-385,148){(a)}
\put(-230,148){(b)}
\put(-70,148){(c)}
\put(-385,-13){(d)}  
\put(-230,-13){(e)}
\put(-70,-13){(f)}  
\caption{Coherence decay signal $\mathcal{L}(t)$ of a NV center in an electron bath of concentration (a)-(c) 0.1 ppm and (d)-(f) 1ppm.
	The different fit results are shown as gray circles. Left column is an exponential fit to $\mathcal{L}(t)$. Middle column is a power fit to the $-\ln(\mathcal{L}(t))$ and the right column is the linear fit on the double logarithmic scale. The red circles indicate the data points included to obtain the fits.}\label{fig:fit_functions}
\end{center}
\end{figure*}

\subsection{Extracting T$_2$ and $p$}
\label{subsec:fit_analysis}

The values of $T_2$ and $p$ obtained for an ensemble average of spin configurations can be quite sensitive to the choice of fitting procedure, specifically if the dynamical the interactions are not sufficiently captured due to the applied approximations in the CCE calculations.
In this section we first discuss the different fitting procedures used so far in literature and show how sensitively the extracted fitting parameters $T_2$ and $p$ are affected by the choice of the fit function. In the second part, we analyze the influence of the time range which is taken into account for the fit.

The first step to obtain the $T_2$  from a decay signal is to fit a function which best describes the noisy signal. There are two main approaches in the literature for extracting $T_2$ and $p$ values from the coherence decay signal $\mathcal{L}(t)$.
Either one fits directly the exponential function, Eq.~(\ref{eq:exp}), to the $\mathcal{L}(t)$ vs.~$t$ data ~\cite{witzel10,wang13,hay20,park22,bauch20,oni20,li21,marcks24}, or one fits a linear function to the $\log(-\ln(\mathcal{L}_t)))$ vs. $\log(t)$ data~\cite{ye19,davis23,hahn24}. Another possibility, which was not yet employed in previous studies, is to fit a power function to the $-\ln(\mathcal{L}_t)$ vs.~$t$ data.
In the following, we refer to these three fitting procedures as \emph{exponential fit}, \emph{linear fit}, and \emph{power fit}, respectively.

In Fig.~\ref{fig:fit_functions} we illustrate the differences in applying these methods to spin-bath concentrations of 0.1 and 1 ppm calculated with the  gCCE2 approximation. Although the data points taken into account (red circles) are the same, the values of $T_2$ and $p$ extracted by the three fit methods are different. The difference between the first two fit methods (the exponential fit and the power fit) are negligible with respect to the numerical uncertainty of the CCE. However, the discrepancy between the linear fit and the other two fits is much more pronounced. 
For instance, the mean values of the changes in T$_2$ and $p$ evaluated from an exponential fit compared to a power fit to the data shown in Fig.~\ref{fig:fit_functions} are 1.6 \% and 3.2 \%, respectively. The outcome of the linear fit leads to a $T_2$ that is, on average, approximately 8.4 \% smaller and a $p$ that is approximately 18 \% larger than the results of the exponential fit. Considering the whole set of bath concentrations studied, we conclude that the exponential fit allows larger $T_2$ and smaller $p$ values while in a linear fit the T$_2$ shrinks and the $p$ grows, in comparison. We point out, that the inclusion of very few data points at the shortest times in the linear fit can notably change its outcome, especially when the set of data is noisy. This is not the case for the other two fitting procedures. In this regard, the choice of an exponential fit function is beneficial to avoid the numerical uncertainties of the noisy data set. 
However, this sensitivity on the choice of the fit function becomes less pronounced for the higher order CCE approximations.

\begin{figure}[]
\begin{center}
{\includegraphics{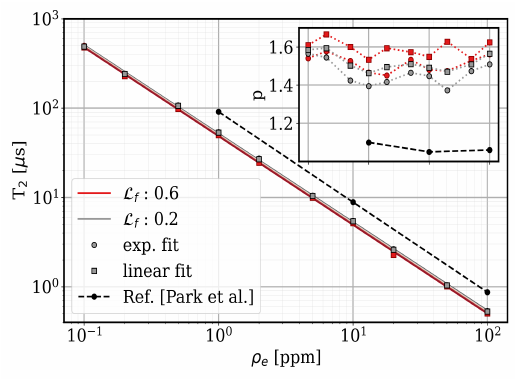}}
\caption{Coherence time T$_2$ obtained from either an exponential (circle symbols) or a linear (square symbols) fit for two different fitting ranges $1>\mathcal{L}(t)>\mathcal{L}_f$, as function of the electron spin bath concentration $\rho_e$. The inset shows the corresponding results for the parameter $p$. The black dashed lines mark the values taken from the CCE2 calculations of Ref.~\cite{park22} for a ``pure electron'' spin bath.}\label{fig:fit_criteria}
\end{center}
\end{figure}

Next, we examine the range of the data points that need to be included for fitting the decay function reliably. Given that in the strongly interacting spin baths the complexity of the spin interactions leads to a more complex noise signal, the fits outlined are not simply applicable to capture the decay process over the full time span. In fact, the $T_2$ and $p$ values obtained from fitting to different time intervals can be very different. 
As already mentioned above, in order to avoid numerical uncertainties, the decay data points at the very early times should be excluded in the linear fit. 

In order to study the influence of the range of coherence decay (and thereby the respective time span) which enters the fitting,
we define a final value $\mathcal{L}_f$ and truncate the data used for fitting accordingly. Here, e.g., $\mathcal{L}_f=0.6$ means that the coherence decay in the range $1>\mathcal{L}(t)>0.6$ is included in the fit.
We determined the values of $T_2$ and $p$ extracted from the gCCE4 data for densities $\rho_e = 0.1$ ppm and 0.2 ppm, and from gCCE3 data for $\rho_e $ = 0.5 -- 100 ppm  using the exponential fit and the linear fit with $\mathcal{L}_f$ varying from 0.6 (i.e.~only the first 40\% of decay are considered) to 0.2 (i.e.~80\% of the decoherence curve is considered). 
The results are plotted in Fig.~\ref{fig:fit_criteria}. 
 
The $T_2$ and $p$ values obtained from an exponential fit (circle symbols) vary on average by 3.5\% and 4.7\%, respectively, when $\mathcal{L}_f$ is varied between 0.6 and 0.2. This variation for the linear fit (square symbols) is slightly larger, on average 7\% and 5.3\% for $T_2$ and $p$, respectively.

We point out that the variation in the extracted $T_2$ and $p$ values obtained from gCCE2 can be as large as 19\% and 22\%, respectively. Therefore, one needs to be cautious using the second order cluster approximation in CCE calculations of strongly interacting spin baths.

The almost linear dependence of $T_2$ on the inverse of the bath-spin concentration (i.e. a linear dependence in the double logarithmic plot with a slope of $\approx-1$) is well captured by our simulations. The sensitivity of the slope of lines in Fig.~\ref{fig:fit_criteria} on the choice of $\mathcal{L}_f$ as well as the choice of fitting method (exponential or linear) is very small. By decreasing $\mathcal{L}_f$ from 0.6 to 0.2 the slope of the exponential fit decreases from -0.995 to -0.991. These values are in a good agreement with the reported values of -1.01~\cite {park22} for the pure electron spin bath and -1.06~\cite{park22}, -1.02~\cite{hahn24} for the P1 center spin bath compared to the experimental values -1.07~\cite{bauch20,li21}.

Furthermore, a general trend in the relation between $T_2$ and $p$ is that the covariance of these two fit parameters is always negative, meaning that these two variables are moving in opposite directions, e.g., when changing $\mathcal{L}_f$. If $T_2$ increases then $p$ decreases and vice versa. This is in line with the decrements of the $p$ values in quasi two-dimensional spin baths reported by 
Refs.~\cite{davis23,hahn24}. Including HF interactions or decreasing the dimensionality of the spin baths prolongs the decay signal. As the consequence of that, the best fit compromise results in a lowest value for $p$.

\section{Summary}\label{sec:summary}

A profound knowledge of the coherence behavior of NV centers in a bath of strongly interacting spins is of high interest for fundamental and applied quantum technology. 
However, theoretical values for the coherence time $T_2$ and the related stretched exponential parameter $p$, as reported in literature, show large variations. We have calculated $T_2$ and $p$ for NV centers in ensembles of unpaired electron spins with concentrations of 0.1--100 ppm and analyzed the potential sources of discrepancy.  
We used the cluster correlation expansion method to calculate the coherence decay. Our study indicates that insufficiencies in the ensemble averaging as well as the choice of parameters determining the numerical convergence of the CCE method may lead to large variations in the resulting coherence parameters. In order to obtain results with 5\% accuracy, at least 250 configurations are required for the ensemble average.  

The results also strongly depend on the procedure used for fitting the time evolution of the averaged coherence decay. We analyzed different fitting approaches to obtain $T_2$ and $p$ values. We found that the results obtained from fitting an exponential function to the numerical decay data may vary up to 10\% compared to the results obtained from the linear fit to the logarithmic data in a double logarithmic plot. 

In order to accurately capture the dynamics of the central NV center spin in strongly interacting electron spin baths, higher order CCE approximations than CCE2 are required. As a compromise between accuracy and numerical workload we have chosen CCE4 for dilute baths and CCE3 for higher bath concentrations.  While the use of CCE2 leads to variations of $\approx$ 20\% in the extracted 
T$_2$ and $p$ values, using the higher order cluster approximations reduces these fit uncertainties below 5\%.

\section{Acknowledgments}
The authors would like to thank P. Sch\"atzle, W. Hahn, D. Maile and J. Stockburger for  fruitful discussions. 
This work was funded by the Ministry of Economic Affairs, Labour and Tourism Baden Württemberg in the context of the Competence Center Quantum Computing Baden-Württemberg (project “SiQuRe II”).

\end{document}